\documentclass[12pt,showpacs]{revtex4}
\usepackage[english]{babel}
\usepackage{graphicx}
\usepackage{epsfig}
\def\beqn{\begin{eqnarray}}
\def\eeqn{\end{eqnarray}}

\def\eqcm{\: ,}           
\renewcommand{\d}{{\rm d}}

\def\be{\begin{equation}}
\def\ee{\end{equation}}
\def\bea{\begin{eqnarray}}
\def\eea{\end{eqnarray}}
\def\ket#1{\hbox{$\vert #1\rangle$}}   
\def\bra#1{\hbox{$\langle #1\vert$}}   
\def\oneh{{\textstyle {1\over 2}}}
\def\onet{{\textstyle {1\over 3}}}
\def\twot{{\textstyle {2\over 3}}}

\def          
\simeq{{\ \lower2pt\hbox{$-$}\mkern-13mu \raise2pt \hbox{$\sim$}\ }}

\def\dsp{\displaystyle}

\begin{document}


\title{Drell-Yan processes, transversity and light-cone wavefunctions}
\author{B.~Pasquini, M.~Pincetti, S.~Boffi
}
\affiliation{Dipartimento di Fisica Nucleare e Teorica, Universit\`a degli
Studi di Pavia and INFN, Sezione di Pavia, Pavia, Italy}
\date{\today}
\begin{abstract}
{The unpolarized, helicity and transversity distributions of quarks in the proton are calculated in the overlap representation of light-cone wavefunctions truncated to the lowest order Fock-space components with three valence quarks. The three distributions at the hadronic scale satisfy an interesting relation consistent with the Soffer inequality. Results are derived in a relativistic quark model including evolution up to the next-to-leading order. Predictions for the double transverse-spin asymmetry in Drell-Yan dilepton production initiated by proton-antiproton collisions are presented. Asymmetries of about 20--30\% are found in the kinematic conditions of the $\cal PAX$ experiment.
}
\end{abstract}
\maketitle 
%


\section{Introduction}

At the parton level the quark structure of the nucleon is described in terms of three quark distributions, namely the quark density $f_1(x)$, the helicity distribution $g_1(x)$ (also indicated $\Delta f(x)$), and the transversity distribution $h_1(x)$ (also indicated $\delta f(x)$). The first two distributions, and particularly  $f_1(x)$, are now well established by experiments in the deep-inelastic scattering (DIS) regime and well understood theoretically as a function of the fraction $x$ of the nucleon longitudinal momentum carried by the active quark~\cite{uno}. Information on the last  leading-twist distribution is missing on the experimental side because $h_1(x)$, being chiral odd, decouples from inclusive DIS and therefore can not be measured in such a traditional source of information. Nevertheless some theoretical activity has been developed in calculating $h_1(x)$ and finding new experimental situations where it can be observed (for a recent review see Ref.~~\cite{Barone}). Among the different proposals the polarized Drell-Yan (DY) dilepton production was recognized for a long time as the cleanest way to access the transversity distribution of quarks in hadrons~\cite{Ralston79,Cortes92,Ji92,JaffeJi}. As a matter of fact, in $pp$ and $p\bar p$ DY collisions with transversely polarized hadrons the leading order (LO) double transverse-spin asymmetry of lepton-pair production involves the product of two transversity distributions, thus giving direct access to them. However, such a measurement is not an easy task because of the technical problems of maintaining the beam polarization through the acceleration. The recently proposed experimental programs at RHIC~\cite{Bunce00} and at GSI~\cite{PAX} have raised renewed interest in theoretical predictions of the double transverse-spin asymmetry in proton-(anti)proton collisions with dilepton production~\cite{EGS04,ABDN04,BCCGR06}.

As reviewed in~\cite{Barone}, $h_1(x)$ has been calculated in a variety of models, including relativistic bag-like, chiral soliton, light-cone, and spectator models. In all these calculations the antiquark transversity is rather small and the $d$-quark distribution turns out to have a much smaller  size than the $u$-quark distribution.

In this paper $h_1(x)$ and the other quark distributions are derived within the framework of the overlap representation of light-cone wavefunctions (LCWFs) originally proposed in Refs.~\cite{diehlbrodsky} to construct generalized parton distributions (GPDs). A Fock-state decomposition of the hadronic state is performed in terms of  $N$-parton Fock states with coefficients representing the momentum LCWF of the $N$ partons. Direct calculation of LCWFs from first principles is a difficult task. On the other hand, constituent quark models (CQMs) have been quite successful in describing the spectrum of hadrons and their low-energy dynamics. At least in the kinematic range where only quark degrees of freedom are effective, it is possible to assume that at the low-energy scale valence quarks can be interpreted as the constituent quarks treated in CQMs.  In the region where they describe emission and reabsorption of a single active quark by the target nucleon, quark GPDs are thus linked to the non-diagonal one-body density matrix in momentum space and can be calculated both in the chiral-even and chiral-odd sector~\cite{BPT03,BPT04,PPB05}. Sea effects represented by the meson cloud can also be integrated into the valence-quark contribution to GPDs~\cite{PB06}. In such an approach the quark distributions, being the forward limit of GPDs, are related to the diagonal part of the one-body density matrix in momentum space.

The paper is organized as follows. In Sect.~\ref{sect:overlap} the overlap representation of LCWFs is briefly reviewed with the aim of linking the parton distributions to CQMs. Results for the three valence quark distributions are discussed at the hadronic scale and after evolution up to the next-to-leading (NLO) in Sect.~\ref{sect:results}. The application to double transverse-spin asymmetry in DY collisions is presented in Sect.~\ref{sect:asymmetry}, and some conclusions are drawn in the final Section.


\section{The overlap representation for parton distributions}
\label{sect:overlap}

In the overlap representation of LCWFs~\cite{diehlbrodsky} the proton wave function with four-momentum $p$ and helicity $\lambda$ is expanded in terms of $N$-parton Fock-space components, i.e.
\be
\label{eq:Fockstate}
\left|p,\lambda\right\rangle = \sum_{N,\beta} 
\int [\d x]_N [\d^2 {\vec k}_\perp]_N\;
\Psi^{[f]}_{\lambda ,N,\beta}(r) \;
\left|N,\beta;k_1,\ldots,k_N \right\rangle \eqcm
\label{eq:representation}
\ee
where $\Psi^{[f]}_{\lambda ,N,\beta}$ is the momentum LCWF of the $N$-parton Fock state $|N,\beta;k_1,\ldots,k_N \rangle$.  The integration measures in Eq.~(\ref{eq:representation}) are defined as
\be
[\d x]_N = \prod_{i=1}^N dx_i \,\delta\left(1-\sum_{i=1}^N x_i\right), \qquad
 [\d^2 {\vec k}_\perp]_N = \frac{1}{(16\pi^3)^{N-1}}\prod_{i=1}^Nd\vec{k}_{\perp,i}\,\delta^2\left(\sum_{i=1}^N{\vec k}_{\perp,i} - {\vec p}_\perp\right),
 \ee
 where $x_i=k_i^+/p^+$ is the fraction of the light-cone momentum of the $i$-th parton and ${\vec k}_{\perp,i}$ its transverse momentum. The argument $r$ of the LCWF represents the set of kinematical variables of the $N$ partons, while  the index $\beta$ labels the quantum numbers of the parton composition and the spin component of each parton. 

Making use of the correct transformation of the wave functions from the (canonical) instant-form to the (light-cone) front-form description,  $\Psi^{[c]}\to\Psi^{[f]}$, and limiting ourselves to the lowest order Fock-space components with three valence quarks, a direct link to wave functions derived in CQMs was established in Refs.~\cite{BPT03,BPT04}. Thus $\Psi^{[f]}_{\lambda ,3,\beta}$ explicitly becomes
\bea
\label{eq:psifc}
 \Psi^{[f]}_{\lambda} (r;\{\lambda_{i}\},\{\tau_i\})
& = &2(2\pi)^3
\left[\frac{1}{M_0}\frac{\omega_1\omega_2\omega_3}
{x_1 x_2 x_3}\right]^{1/2}
\sum_{\mu_1\mu_2\mu_3}{D}^{1/2\,*}_{\mu_1\lambda_1}(R_{cf}(k_1))
{D}^{1/2\,*}_{\mu_2\lambda_2}(R_{cf}(k_2))
\nonumber\\
& & \qquad\times {D}^{1/2\,*}_{\mu_3\lambda_3}(R_{cf}(k_3))
\,
\Psi_\lambda^{[c]}(\{\vec{k}_i\};\{\mu_i\},\{\tau_i\}),
\eea
where $M_0=\omega_1+\omega_2+\omega_3$ is the mass of the noninteracting three-quark system, with $\omega_i\equiv k^0_i = (k^+_i + k^-_i)/\sqrt{2}$, and the matrices ${D}^{1/2}_{\mu_i\lambda_i}(R_{cf}(k_i))$ are given by the spin-space representation of the Melosh rotation $R_{cf}$,
\bea
{D}^{1/2}_{\lambda\mu}(R_{cf}(k)) & = &
\bra{\lambda}R_{cf}(x M_0,\vec{k}_\perp)\ket{\mu} \nonumber\\
& = & \bra{\lambda}
\frac{m+ xM_0-i\vec{\sigma}\cdot(\hat{\vec{z}}\times\vec{k}_\perp)}
{\sqrt{(m+ xM_0)^2+\vec{k}_\perp^2}}\ket{\mu}.
\eea

In this approach the ordinary (unpolarized) parton distributions of flavor $q$~\cite{BL} can be recovered taking into account that in this case the Melosh rotation matrices combine to the identity matrix:
\be
f_1^q(x) = 
\sum_{\lambda_i\tau_i}\sum_{j=1}^3 \delta_{\tau_j\tau_q}\, 
 \int[d x]_3[d\vec{k}_\perp]_3\, \delta(x-x_j) 
\vert\Psi_\lambda^{[f]}(\{x_i\},\{\vec{k}_{\perp,i}\};\{\lambda_i\},\{\tau_i\})\vert^2,
\label{eq:parton}
\ee
where the helicity $\lambda$ of the nucleon can equivalently be taken positive or negative.
Analogously, the following simple expressions are obtained for the polarized quark distribution of flavor $q$~\cite{BPT04}
\be
g_1^q(x) =
\sum_{\lambda_i\tau_i}\sum_{j=1}^3 \delta_{\tau_j\tau_q}\,
\mbox{sign}\,(\lambda_j)\,  \int[d x]_3[d\vec{k}_\perp]_3\, \delta(x-x_j) 
\vert\Psi_+^{[f]}(\{x_i\},\{\vec{k}_{\perp,i}\};\{\lambda_i\},\{\tau_i\})\vert^2,
\label{eq:helicity}
\ee
and for the quark transversity distributions $h_1^q(x)$~\cite{PPB05}:
 \be
\label{eq:forward}
h^{q}_1(x)
= \sum_{\lambda^t_i\,\tau_i}\sum_{j=1}^3 \delta_{\tau_j\tau_q}\,
\mbox{sign}\,(\lambda^{t}_j)
\int[dx]_3[d\vec{k}_\perp]_3\, \delta(x-x_j) 
\vert\Psi_{\uparrow}^{[f]}(\{x_i\},\{\vec{k}_{\perp,i}\};\{\lambda^t_i\},\{\tau_i\})
\vert^2,
\label{eq:transversity}
\ee
where $\lambda^t_i$ is the transverse-spin component of the quark and, as usual,
 the transversity basis for the nucleon spin states is obtained from the helicity basis as follows:
\be
\ket{p,\uparrow}=\frac{1}{\sqrt{2}}(|p,+\rangle+|p,-\rangle), \qquad
\ket{p,\downarrow}=\frac{1}{\sqrt{2}}(|p,+\rangle -|p,-\rangle).
\label{eq:trans_spinor}
\ee

Expressions~(\ref{eq:parton}), (\ref{eq:helicity}) and (\ref{eq:transversity}) exhibit the well known probabilistic content of parton distributions. Eq.~(\ref{eq:parton}) gives the probability of finding a quark with a fraction $x$ of the longitudinal momentum of the parent nucleon, irrespective of its spin orientation. The helicity distribution $g_1^q(x)$ in Eq.~(\ref{eq:helicity}) is the number density of quarks with helicity $+$ minus the number density  of quarks with helicity $-$, assuming the parent nucleon to have helicity $+$. The transversity distribution $h^{q}_1(x)$ in Eq.~(\ref{eq:transversity}) is the number density of quarks with transverse polarization $\uparrow$ minus the number density  of quarks with transverse polarization $\downarrow$, assuming the parent nucleon to have transverse polarization $\uparrow$. 

In the instant form it is convenient to separate the spin-isospin component from the space part of the proton wave function and to assume SU(6) symmetry, i.e.
\be
\label{eq:separated}
\Psi^{[c]}_\lambda (\{\vec{k}_i\},\{\lambda_i\},\{\tau_i\})
= \psi(\vec{k}_1,\vec{k}_2,\vec{k}_3)
\Phi_{\lambda\tau}(\lambda_1,\lambda_2,\lambda_3,\tau_1,\tau_2,\tau_3) ,
\ee
where
\bea
& & \Phi_{\lambda\tau}(\lambda_1,\lambda_2,\lambda_3,\tau_1,\tau_2,\tau_3)
\nonumber\\ 
& & \quad=
\frac{1}{\sqrt{2}}\left[\Phi^0_\lambda(\lambda_1,\lambda_2,\lambda_3)
\Phi^0_\tau(\tau_1,\tau_2,\tau_3) +
\Phi^1_\lambda(\lambda_1,\lambda_2,\lambda_3)\Phi^1_\tau(\tau_1,\tau_2,\tau_3)
\right],
\label{eq:spinisospin}
\eea
with the superscripts $0$ and $1$ referring to the total spin or isospin of the pair of quarks 1 and 2. Thus
we find
 \begin{eqnarray}
\label{eq:f1}
f^q_1(x)&=&
 \left(2\delta_{\tau_q 1/2}+\delta_{\tau_q -1/2}\right)
\int[dx]_3[d\vec{k}_\perp]_3\, \delta(x-x_3) 
\vert \psi(\{x_i\},\{\vec{k}_{\perp,i}\})\vert^2,
\\
\label{eq:g1}
g^q_1(x)&=&\left(\frac{4}{3}\delta_{\tau_q 1/2}
-\frac{1}{3}\delta_{\tau_q -1/2}\right)
\int[dx]_3[d\vec{k}_\perp]_3\, \delta(x-x_3) 
\vert \psi(\{x_i\},\{\vec{k}_{\perp,i}\})\vert^2
\mathcal{M},\\
\label{eq:h1}
h^q_1(x)&=&\left(\frac{4}{3}\delta_{\tau_q 1/2}
-\frac{1}{3}\delta_{\tau_q -1/2}\right)
\int[dx]_3[d\vec{k}_\perp]_3\, \delta(x-x_3) 
\vert \psi(\{x_i\},\{\vec{k}_{\perp,i}\})\vert^2
\mathcal{M}_T,
\end{eqnarray}
where~\cite{PPB05,MaSS}
\bea
\mathcal{M}&=& \frac{(m+ x_3M_0)^2 - \vec{k}^2_{\perp,3}}{(m+ x_3M_0)^2 + \vec{k}^2_{\perp,3}},
\label{eq:m}\\
& &\nonumber\\
\mathcal{M}_T&=& \frac{(m+ x_3M_0)^2}{(m+ x_3M_0)^2 + \vec{k}^2_{\perp,3}},
\label{eq:mt}
\eea
and the expectation values on the normalized nucleon momentum wavefunction of the contribution coming from Melosh rotations satisfy
\be
2\langle \mathcal{M}_T\rangle = \langle \mathcal{M}\rangle +1.
\ee
Therefore the following relations hold
\begin{eqnarray}
\label{eq:sofferbis}
h^u_1(x)=\frac{1}{2}g^u_1(x)+\frac{1}{3}f^u_1(x),
& \quad&
h^d_1(x)=\frac{1}{2}g^d_1(x)-\frac{1}{6}f^d_1(x),
\end{eqnarray}
which are compatible with the Soffer inequality~\cite{soffer}:
\be
\vert h^q_1(x)\vert \le \frac{1}{2}[f_1^q(x)+g_1^q(x)].
\label{eq:soffer}
\ee

In the nonrelativistic limit, corresponding to $\vec{k}_\perp=0$, i.e. $\mathcal{M}_T=\mathcal{M}=1$, one obtains  $h^u_1=g^u_1=\twot f^u_1$ and $h^d_1=g^d_1=-\onet f^d_1$ as expected from general principles~\cite{JaffeJi}.


\section{Results}
\label{sect:results}

As an application of the general formalism reviewed in the previous section we consider the valence-quark contribution to the parton distributions starting from an instant-form SU(6) symmetric wave function of the proton, Eqs. (\ref{eq:separated}) and (\ref{eq:spinisospin}), derived in the relativistic quark model of Ref.~\cite{Schlumpf94a}.  In particular, we use the Lorentzian shape wavefunction of Ref.~\cite{Schlumpf94a} with parameters fitted to the magnetic moments of the proton and the neutron and the axial-vector coupling constant $G_A$ and giving also a good agreement with the experimental nucleon electroweak form factors in a large $Q^2$ range. Furthermore, we note that SU(6) symmetry is broken in the LCWF $\Psi^{[f]}$~\cite{Cardarelli} as a consequence of the transformation~(\ref{eq:psifc}).

The distributions in Eqs.~(\ref{eq:f1}), (\ref{eq:g1}) and (\ref{eq:h1}) are defined at the hadronic scale $Q_0^2$ of the model. In order to make predictions for experiments, a complete knowledge of the evolution up to NLO is indispensable. According to Ref.~\cite{JaffeRoss} we assume that twist-two matrix elements calculated at some low scale in a quark model can be used in conjunction with QCD perturbation theory. Starting from a scale where the long-range (confining) part of the interaction  is dominant, we generate the perturbative contribution by evolution at higher scale. In the case of transversity the Dokshitzer-Gribov-Lipatov-Altarelli-Parisi (DGLAP) $Q^2$ evolution equation~\cite{DGLAP} is simple. In fact, being chirally odd, the quark transversity distributions do not mix with the gluon distribution and therefore the evolution is of the non-singlet type. The leading order (LO) anomalous dimensions were first calculated in Ref.~\cite{Baldracchini81} but promptly forgotten. They were recalculated by Artru and Mekhfi~\cite{Artru90}. The one-loop coefficient functions for Drell-Yan processes are known in different renormalization schemes~\cite{Vogelsang93,CKM94,Kamal96}. The NLO (two-loop) anomalous dimensions were also calculated in the Feynman gauge in Refs.~\cite{Hayashigaki97,Kumano97}  and in the light-cone gauge~\cite{Vogelsang98}. The two-loop splitting functions for the evolution of the transversity distribution were calculated in Ref.~\cite{Vogelsang98}. The LO DGLAP $Q^2$ evolution equation for the transversity distribution $h_1(x)$ was derived in Ref.~\cite{Artru90} and its numerical analysis is discussed in Refs.~\cite{Barone97,Barone97b}. 

A numerical solution of the DGLAP equation for the transversity distribution $h_1(x)$ was given at LO and NLO in Refs.~\cite{HKM96b,CC04}.  In Ref.~\cite{HKM96b} the DGLAP integrodifferential equation is solved in the variable $Q^2$ with the Euler method replacing the Simpson method previously used in the cases of unpolarized~\cite{Miyama96} and longitudinally polarized~\cite{HKM96a}  structure functions. 

In the present analysis the FORTRAN code of Ref.~\cite{HKM96b} has been applied within the $\overline{MS}$ renormalization scheme and the input distributions calculated at the hadronic scale according to the model explained in Sect.~\ref{sect:overlap} were evolved up to NLO. The model scale $Q_0^2=0.079$ GeV$^2$ was determined by matching the value of the momentum fraction carried by the valence quarks, as computed in the model, with that obtained by evolving backward the value experimentally determined at large $Q^2$. The strong coupling $\alpha_s(Q^2)$ entering the code at NLO is computed by solving the NLO transcendental equation numerically,
\be
\ln {Q^2\over\Lambda_{\rm NLO}^2}-{4\,\pi\over\beta_0\,\alpha_s} + 
{\beta_1\over\beta_0^2}\,\ln\left[
{4\,\pi\over\beta_0\,\alpha_s} + {\beta_1\over\beta_0^2}\right] = 0\,,
\label{0:10}
\ee
as obtained from the renormalization group analysis~\cite{alpha}. It differs from the more familiar expression used in Ref.~\cite{HKM96b},
\begin{equation}
{\alpha_s(Q^2) \over 4\pi}={1 \over \beta_0\ln(Q^2/\Lambda_{\rm NLO}^2)}
\left(1-{\beta_1 \over \beta_0^2}\,{\ln\ln(Q^2/\Lambda_{\rm NLO}^2)\over
\ln(Q^2/\Lambda_{\rm NLO}^2)}\right),
\label{1:10}
\end{equation}
valid only in the limit $Q^2\gg\Lambda_{\rm NLO}^2$, where $\Lambda_{\rm NLO}$ is the so-called QCD scale parameter.


\begin{figure}[ht]
\begin{center}
\epsfig{file=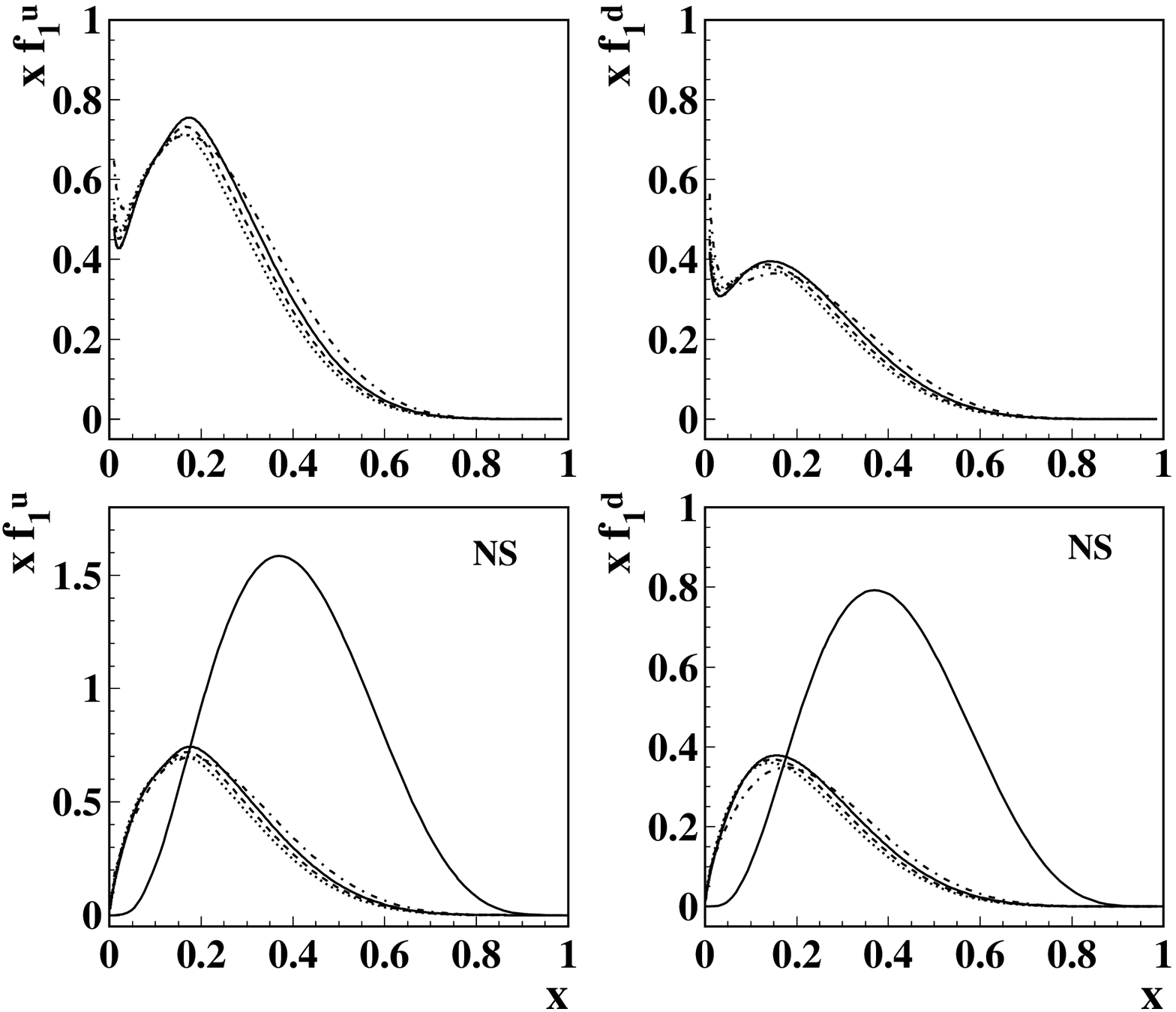,  width=32 pc}
\end{center}
\vspace{-0.4cm}
\caption{Evolution of the parton distribution for the $u$ (left panel) and $d$ (right panel) quark. In the lower panels starting from the hadronic scale $Q_0^2=0.079$ GeV$^2$ (upper curve), LO non-singlet distributions are shown at different scales ($Q^2=5$ GeV$^2$, solid lines; $Q^2=9$ GeV$^2$, dashed lines; $Q^2=16$ GeV$^2$, dotted lines) together with NLO distributions at $Q^2=5$ GeV$^2$ (dot-dashed lines). LO and NLO total distributions are shown in the upper panels with the same line convention. The parametrization of Ref.~\cite{GRV} NLO evolved at 5 GeV$^2$ is also shown by small stars.}
\label{fig:fig1}
\end{figure}


\begin{figure}
\begin{center}
\epsfig{file=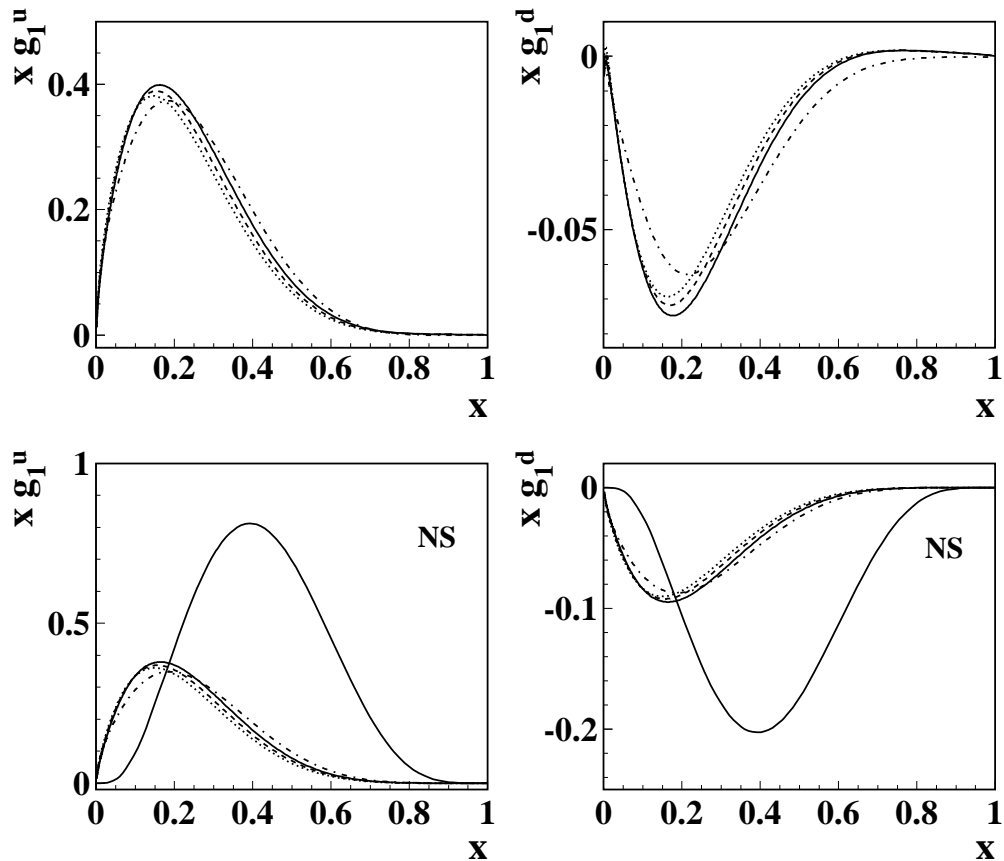,  width=32 pc}
\end{center}
\vspace{-0.4cm}
\caption{The same as in Fig.~\ref{fig:fig1}, but for the helicity distribution.}
\label{fig:fig2}
\end{figure}


\begin{figure}
\begin{center}
\epsfig{file=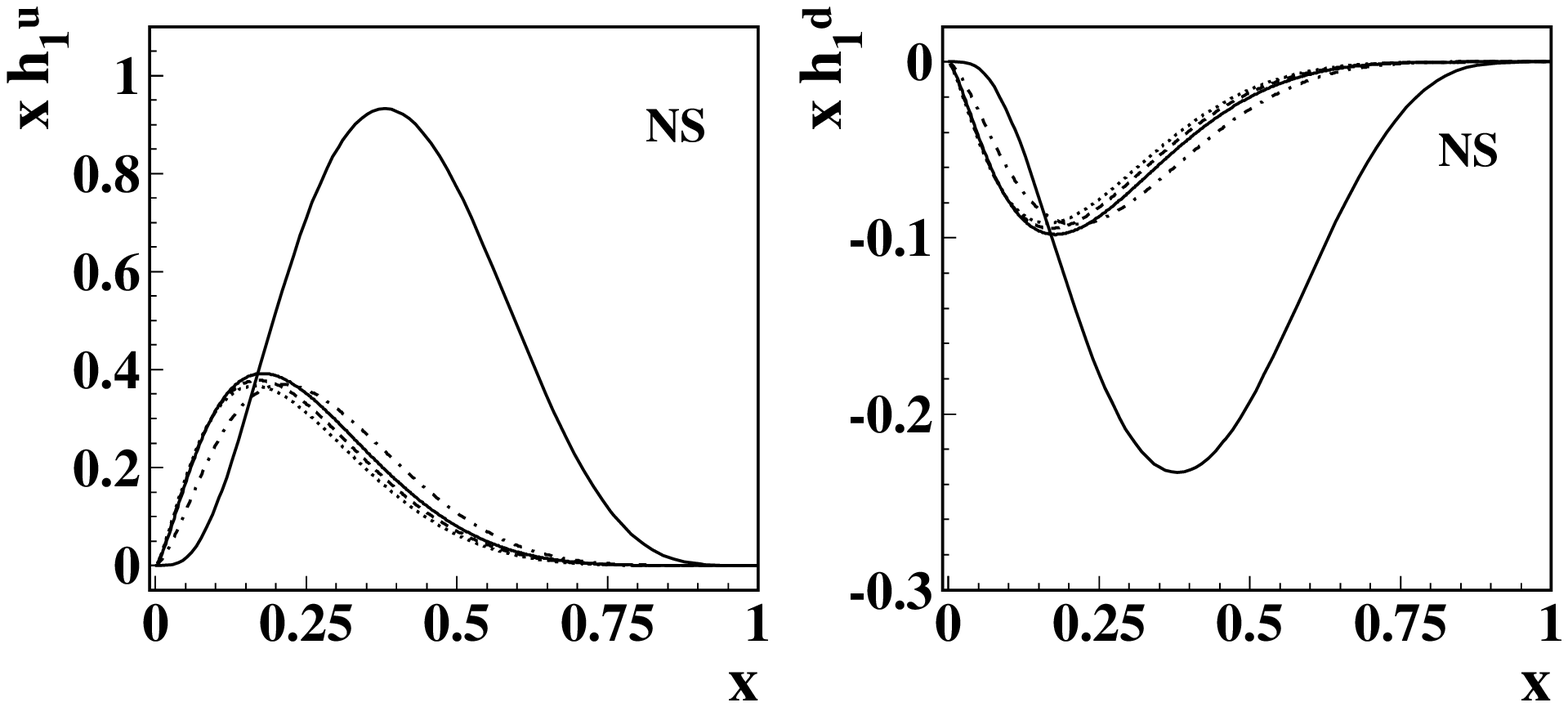,  width=32 pc}
\end{center}
\vspace{-5cm}
\caption{Evolution of the transversity distribution for the $u$ (left panel) and $d$ (right panel) quark. Starting from the hadronic scale $Q_0^2=0.079$ GeV$^2$ (upper curve), LO non-singlet distributions are shown at different scales ($Q^2=5$ GeV$^2$, solid lines; $Q^2=9$ GeV$^2$, dashed lines; $Q^2=16$ GeV$^2$, dotted lines) together with NLO distributions at $Q^2=5$ GeV$^2$ (dot-dashed lines).}
\label{fig:fig3}
\end{figure}


\begin{figure}
\begin{center}
\epsfig{file=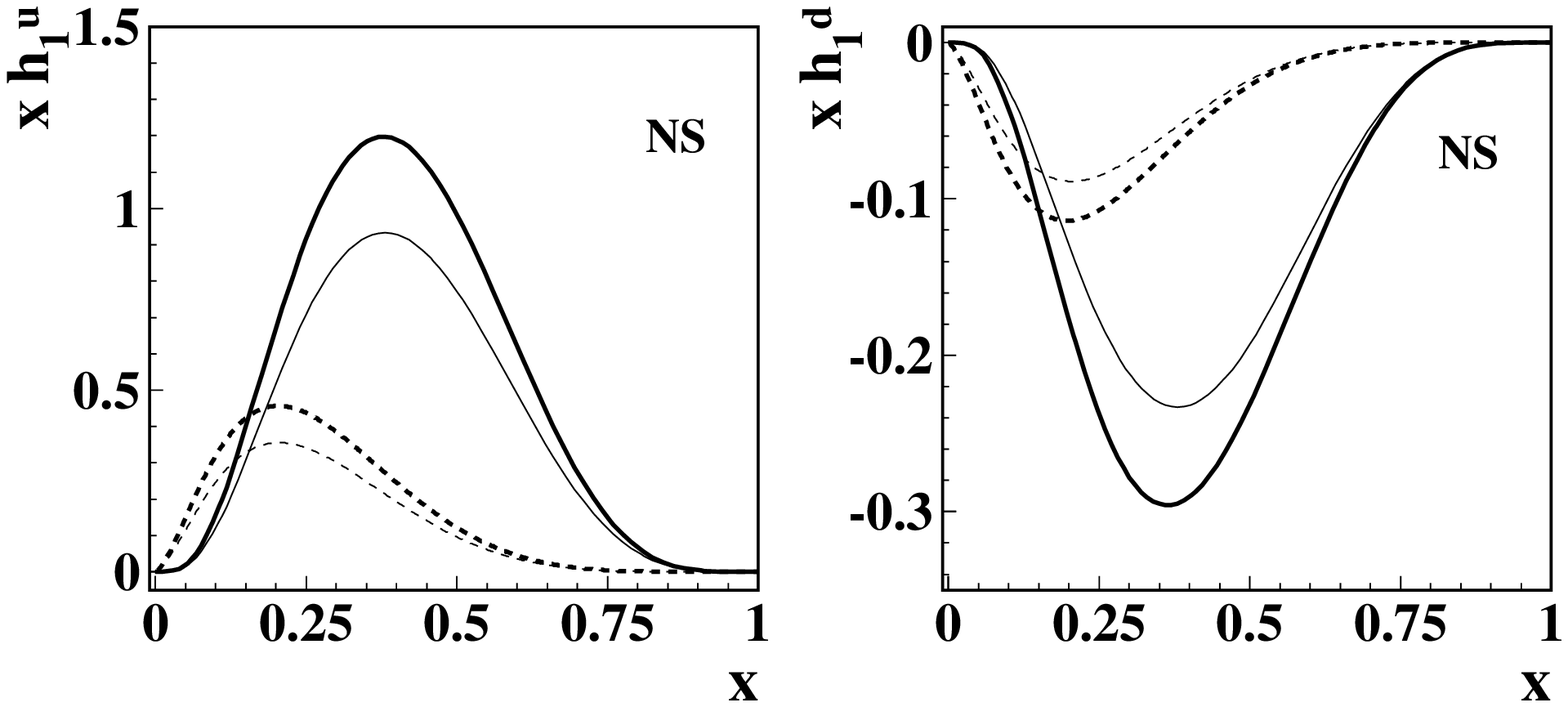,  width=32 pc}
\end{center}
\vspace{-5cm}
\caption{The transversity distribution obtained with the LCWFs of the present model (thin lines) compared with the Soffer bound, Eq.~(\ref{eq:saturate}), (thick lines) for the $u$ (left panel) and $d$ (right panel) quark . Solid lines for the results at the hadronic scale $Q_0^2=0.079$ GeV$^2$, the dashed lines obtained by NLO evolution at $Q^2=9$ GeV$^2$, respectively. }
\label{fig:fig4}
\end{figure}


Together with the input distributions at the hadronic scale the non-singlet (valence) contribution of the three parton distributions is shown in Figs.~\ref{fig:fig1} to~\ref{fig:fig3} at LO and NLO at different scales of $Q^2$. In the case of the unpolarized and polarized distributions, Figs.~\ref{fig:fig1} and \ref{fig:fig2} respectively, the result of evolution of the total distributions is also presented. Quite generally, the $Q^2$ dependence of the evolution is weak within a given order, while small effects are introduced when going from LO to NLO, as exemplified by the dot-dashed curves at $Q^2=5$ GeV$^2$ in Figs.~\ref{fig:fig1} to~\ref{fig:fig3}. Thus, convergence of the perturbative expansion is very fast and one can safely limit himself to LO. 

The size of the $d$-quark distributions is always smaller than that of the $u$-quark distribution, particularly in the case of transversity, confirming results obtained with other models (see, e.g., \cite{EGS04}).

Taking into account that the model at the hadronic scale only considers valence quarks and the sea is only generated perturbatively, the overall behavior of $f_1(x)$ is in reasonable agreement with available parametrizations~\cite{GRV}. One may notice the faster fall-off of the tail of $f_1^u(x)$ at large $x$ in our model with respect to the parametrization~\cite{GRV} that will have some consequences in the predicted double transverse-spin asymmetry in Sect. IV. 
As for $g_1(x)$, the missing sea and gluon contributions are crucial to compare our model results with the available parametrizations~\cite{GRSV}. Therefore, $g_1(x)$ is shown here for completeness, but it requires a more systematic study (e.g. along the lines of Ref.~\cite{PB06}) that goes beyond the goal of the present investigation focused on the double transverse-spin asymmetry. 

However, comparison of $h_1(x)$ and $g_1(x)$ is here legitimate because $h_1(x)$ is determined by valence contributions, as it is $g_1(x)$  in our model. As can be see in Figs.~\ref{fig:fig2} and \ref{fig:fig3} they are rather different not only after evolution, but especially at the hadronic scale of the model. This contrasts with the popular guess $h_1(x)\approx g_1(x)$ motivated on the basis of the nonrelativistic quark model.

In any case the Soffer inequality~(\ref{eq:soffer}) at each order is always satisfied by the three quark distributions calculated with the LCWFs of the present model (see Fig.~\ref{fig:fig4}). In contrast, saturation of the Soffer bound, i.e. assuming
\be
\vert h_1^q(x)\vert =\frac{1}{2}\left[ f_1^q(x)+ g_1^q(x)\right],
\label{eq:saturate}
\ee
is neither reached at the hadronic scale of the model nor is it a conserved property during evolution. In fact, starting at the hadronic scale with the transversity distribution given by Eq.~(\ref{eq:saturate}), the result of LO and NLO evolution diverges from that obtained when calculating the transversity according to Eq.~(\ref{eq:saturate}) after separate evolution of $f_1$ and $g_1$. Since the two sides of Eq.~(\ref{eq:saturate}) give different results under evolution, in model calculations the choice of the initial hadronic scale is crucial. This fact should put some caution about the possibility of making predictions with the transversity distribution guessed from $f_1$ and $g_1$ as, e.g., in the case of the double transverse-spin asymmetry in DY processes (see Refs.~\cite{ABDN04,BCCGR06} and Fig.~\ref{fig:fig6} below).

A similar situation occurs when the transversity distribution is derived from $f_1$ and $g_1$ according to the relations~(\ref{eq:sofferbis}), with the difference that these relations are exact at the hadronic scale when only valence quarks are involved.


\section{The double transverse-spin asymmetry}
\label{sect:asymmetry}

In order to directly access transversity via Drell-Yan lepton pair production one has to measure the double transverse-spin asymmetry $A_{TT}$ in collisions between two transversely polarized hadrons:
\be
A_{TT} = \frac{d\sigma^{\uparrow\uparrow} - d\sigma^{\uparrow\downarrow}}{d\sigma^{\uparrow\uparrow} + d\sigma^{\uparrow\downarrow}},
\ee
with the arrows denoting the transverse directions along which the two colliding hadrons are polarized. 

At LO, i.e. considering only the quark-antiquark annihilation graph, the double trans\-verse-spin asymmetry for the process $p^\uparrow p^\uparrow\to\ell^+\ell^- X$ mediated by a virtual photon is given by
\be
A^{pp}_{TT} = a_{TT}\,\frac{\dsp \sum_q e_q^2\left[h_1^q(x_1,Q^2)h_1^{\bar q}(x_2,Q^2) + (1\leftrightarrow 2)\right]}
{\dsp \sum_q e_q^2\left[f_1^q(x_1,Q^2)f_1^{\bar q}(x_2,Q^2) + (1\leftrightarrow 2)\right]},
\label{eq:ppasym}
\ee
where $e_q$ is the quark charge, $Q^2$ the invariant mass square of the lepton pair (dimuon), and  $x_1x_2=Q^2/s$ where $s$ is the Mandelstam variable. The quantity $a_{TT}$ is the spin asymmetry of the QED elementary process $q\bar q\to \ell^+ \ell^-$, i.e.
\be
a_{TT}(\theta,\phi) = \frac{\sin^2\theta}{1+\cos^2\theta}\cos(2\phi),
\ee
with $\theta$ being the production angle in the rest frame of the lepton pair and $\phi$ the angle between the dilepton direction and the plane defined by the collision and polarization axes.

After the first simple encouraging estimates~\cite{Ji92}, some phenomenological studies of DY dimuon production at RHIC have been presented~\cite{Vogelsang93,CKM94,BouS94,BS95,Barone97b,Barone97a,MSSV99} indicating that accessing transversity is very difficult under the kinematic conditions of the proposed experiments with $pp$ collisions~\cite{Bunce00}. The main reason is that $A^{pp}_{TT}$ in Eq.~(\ref{eq:ppasym}) involves the product of quark and antiquark transversity distributions. The latter are small in a proton, even if they were as large as to saturate the Soffer inequality; moreover, the QCD evolution of transversity is such that, in the kinematical regions of RHIC data, $h_1(x,Q^2)$ is much smaller than the corresponding values of $g_1(x,Q^2)$ and $f_1(x,Q^2)$. This makes the measurable $A^{pp}_{TT}$ at RHIC very small, no more than a few percents~\cite{Barone97b,Barone97a,MSSV99}.

A more favorable situation is expected by using an antiproton beam instead of a proton beam~\cite{PAX,EGS04,ABDN04,BCCGR06,BR05}. In $p\bar p$ DY the LO asymmetry $A^{p\bar p}_{TT}$ is proportional to a product of quark transversity distributions from the proton and antiquark distributions from the antiproton which are connected by charge conjugation, e.g.
\be
h_1^{u/p} (x) = h_1^{\bar u/\bar p} (x).
\ee 
Therefore one obtains
\be
A^{p\bar p}_{TT} = a_{TT}\frac{\dsp \sum_q e_q^2\left[h_1^q(x_1,Q^2)h_1^q(x_2,Q^2) + h_1^{\bar q}(x_1,Q^2)h_1^{\bar q}(x_2,Q^2)\right]}
{\dsp \sum_q e_q^2\left[f_1^q(x_1,Q^2)f_1^q(x_2,Q^2) + f_1^{\bar q}(x_1,Q^2)f_1^{\bar q}(x_2,Q^2))\right]},
\label{eq:asy-lo}
\ee
so that in this case the asymmetry is only due to valence quark distributions. 

Quantitative estimates of $A^{p\bar p}_{TT}$ for the kinematics of the proposed $\cal PAX$ experiment at GSI~\cite{PAX} were presented in Refs.~\cite{EGS04,ABDN04,BCCGR06}. On the basis of predictions from the chiral quark-soliton model~\cite{EGS04}, the LO DY asymmetries turn out to be large, of the order of 50\%, increasing with $Q^2$ and almost entirely due to $u$-quarks. In contrast, they are in the range 20--40\% in a phenomenological analysis~\cite{ABDN04,BCCGR06} where $A^{p\bar p}_{TT}$ is appropriately evolved at NLO starting from two extreme possibilities at some typical low scale $\mu_0\le 1$ GeV. One assumption was $h_1(x)=g_1(x)$, as in the nonrelativistic case. The second ansatz for the transversity was the saturation of Soffer's inequality according to Eq.~(\ref{eq:saturate}). The two possibilities have been considered to give a lower and upper bound for the transversity and, consequently, for the $A^{p\bar p}_{TT}$ asymmetry.

NLO effects hardly modify the asymmetry since the $K$ factors of the transversely polarized and unpolarized cross sections are similar to each other and therefore almost cancel out in the ratio~\cite{Shimizu}. In addition, NLO effects are rather small on the quark distributions obtained in Sect.~\ref{sect:results} (see Figs.~\ref{fig:fig1}--\ref{fig:fig3}). Therefore, the following discussion is limited to LO.


\begin{figure}
\begin{center}
\epsfig{file=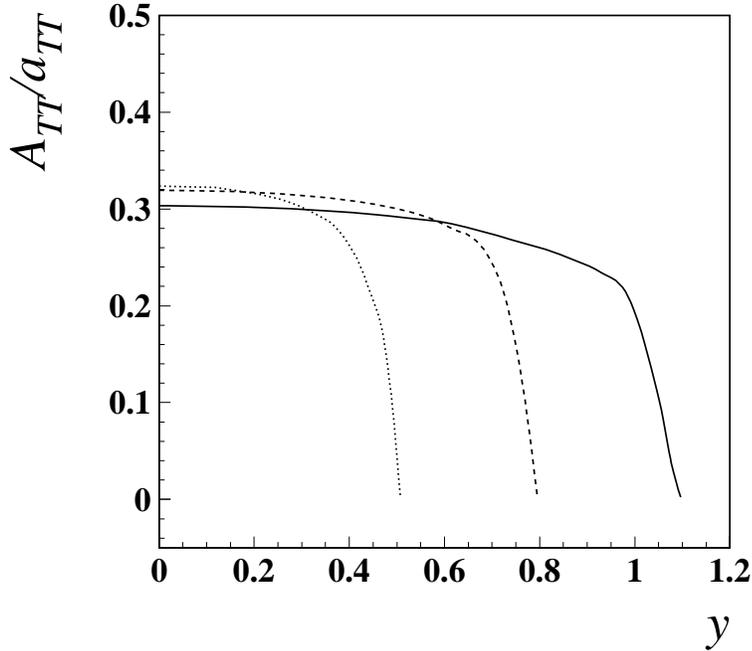,  width=30 pc}
\end{center}
\vspace{-0.4cm}
\caption{The double transverse-spin asymmetry $A^{p\bar p}_{TT}/a_{TT}$ calculated with the parton distributions of the present model as a function of the rapidity $y$ at different scales: $Q^2=5$ GeV$^2$, solid line; $Q^2=9$ GeV$^2$, dashed line; $Q^2=16$ GeV$^2$, dotted line.}
\label{fig:fig5bis}
\end{figure}


Using the unpolarized quark  and transversity distributions derived in Sect.~\ref{sect:overlap}, results for $s=45$ GeV$^2$ and different values of $Q^2$ are plotted in Fig.~\ref{fig:fig5bis} in terms of the rapidity
\be
y=\frac{1}{2}\ln\frac{x_1}{x_2}.
\ee
An asymmetry of about 30\% (comparable with Refs.~\cite{ABDN04,BCCGR06}) is obtained, with a $Q^2$ dependence in agreement with Ref.~\cite{EGS04}.

This result confirms the possibility of measuring the double transverse-spin asymmetry under conditions that will be probed by the proposed $\cal PAX$ experiment. In such conditions, assuming the LO expression (\ref{eq:asy-lo}) for the observed asymmetry one could gain direct information on the transversity distribution 
following previous analyses~\cite{EGS04,ABDN04,BCCGR06}, where the quark densities $f_1^{q,\bar q}(x,Q^2)$ 
are taken from the GRV98 parametrizations~\cite{GRV}.
The resulting transversity distributions could be compared with model predictions.


\begin{figure}
\begin{center}
\epsfig{file=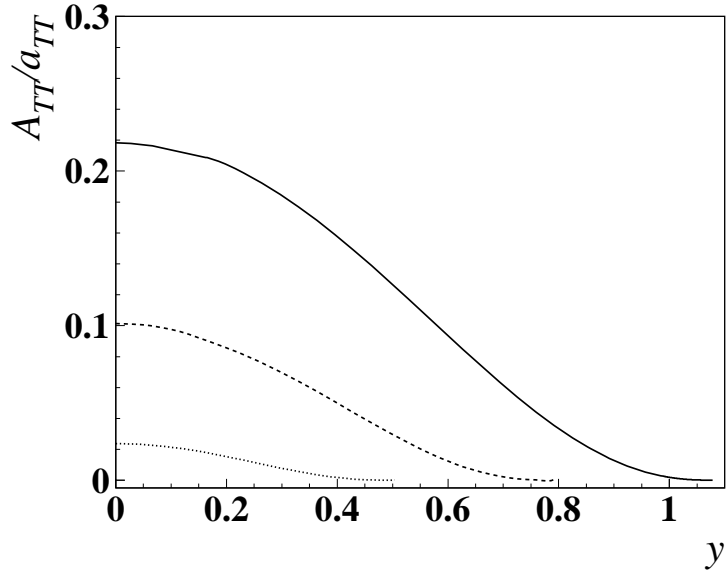,  width=24 pc}
\end{center}
\vspace{-0.4cm}
\caption{The same as in Fig.~\ref{fig:fig5bis} but assuming the GRV98~\cite{GRV} quark density.}
\label{fig:fig5}
\end{figure}



\begin{figure}
\begin{center}
\epsfig{file=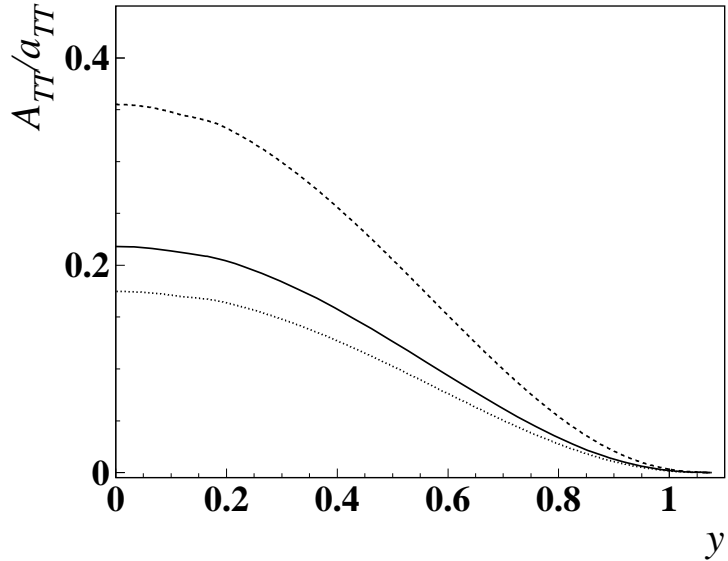,  width=24 pc}
\end{center}
\vspace{-0.4cm}
\caption{The double transverse-spin asymmetry $A^{p\bar p}_{TT}/a_{TT}$ as a function of the rapidity $y$ at $Q^2=5$ GeV$^2$ and $s=45$ GeV$^2$. Solid curve: calculation with $h_1$ obtained with the LCWFs of the present model. Dashed curve: calculation with an input $h_1= \oneh(g_1+f_1)$. Dotted curve: calculation with an input $h_1=g_1$.}
\label{fig:fig6}
\end{figure}


According to this strategy, with the present model the antiquark distributions $h_1^{\bar q}(x,Q^2)$ are identically vanishing and $h_1^q(x,Q^2)$ contains only valence quark contributions. Assuming a negligible sea-quark contribution the corresponding asymmetry would thus give direct access to $h_1^q(x,Q^2)$ and would look like that shown in Fig.~\ref{fig:fig5}.
The results
indicate a strong $Q^2$ dependence suggesting moderate values of $Q^2$, e.g. $Q^2=5$ to $10$ GeV$^2$, in order to have an appreciable asymmetry of about 10--20\% at the proposed $\cal PAX$ experiment at GSI~\cite{PAX}. It is remarkable that, contrary to the result of Ref.~\cite{EGS04}, in the present model $Q^2$ evolution produces a decreasing LO asymmetry with increasing $Q^2$ as a consequence of the opposite $Q^2$ dependence of the theoretical $h_1$ and the phenomenological  $f_1$. In fact, in the range of $x$-values explored by the chosen kinematic conditions ($x\ge0.3$) $h_1$ with its valence quark contribution has a larger fall-off with $Q^2$ than the  GRV98 $f_1$ as shown in Fig.~\ref{fig:fig1}. Furthermore, one may notice that with the present model a much lower asymmetry is predicted than with the chiral quark-soliton model~\cite{EGS04} and even lower than the phenomenological analysis of Refs.~\cite{ABDN04,BCCGR06}. 

In general, one can anticipate upper and lower limits for the theoretical asymmetry depending on the  upper and lower bounds that the transversity has to satisfy. The saturated Soffer bound (\ref{eq:saturate}), i.e. $h_1= \oneh(g_1+f_1)$, represents the upper bound of $h_1$ at any scale. The lower bound is given by the nonrelativistic approximation $h_1=g_1$. At the hadronic scale the transversity calculated with any LCWFs including valence quarks only should have intermediate values satisfying the conditions in Eq.~(\ref{eq:sofferbis}). Under evolution Eq.~(\ref{eq:sofferbis}) does no longer hold, but still the evolved transversity has to lie in between the correspondingly evolved upper and lower bounds. Assuming the LCWFs of the present model, the same asymmetry shown in Fig.~\ref{fig:fig5} at $Q^2=5$ GeV$^2$ is compared in Fig.~\ref{fig:fig6} 
with the asymmetry calculated when the transversity is evolved starting from an input  at the hadronic scale either given by the saturated Soffer bound $h_1= \oneh(g_1+f_1)$ (dashed curve) or assuming the nonrelativistic approximation $h_1=g_1$ (dotted curve), with $f_1$ and $g_1$ calculated in the present model. 
The difference between the dotted and solid curves gives an estimate of the relativistic effects in the calculation of $h_1$. On the other side, the model calculation with an input $h_1$ satisfying Eq.~(\ref{eq:sofferbis}) leads to an asymmetry much lower than in the case of the saturated Soffer bound.


\section{Conclusions}

\bigskip

The unpolarized, helicity and transversity distributions of quarks in the proton are calculated in the overlap representation of light-cone wavefunctions truncated to the lowest order Fock-space components with three valence quarks. The light-cone wavefunctions have been defined making use of the correct covariant connection with  the instant-form wavefunctions used in any constituent quark model. The quark distributions have been evolved to leading order and next-to-leading order of the perturbative expansion with the remarkable result that NLO effects are rather small compared to LO. The three distributions at the hadronic scale satisfy an interesting relation consistent with the Soffer inequality. In particular, the transversity distribution has been used to predict the double transverse-spin asymmetry in dilepton production with Drell-Yan collisions between transversely polarized beams of protons and antiprotons. As a function of rapidity the asymmetry calculated in the model is about 30\% for $s=45$ GeV$^2$, slightly increasing with $Q^2$. In contrast, when using phenomenological unpolarized quark distributions together with the transversity distribution derived in the present model,  the asymmetry turns out to be smaller than previous predictions, e.g. about 10--20\% at $Q^2=5$ GeV$^2$ and $s=45$ GeV$^2$, and rapidly decreases with increasing $Q^2$. This is due to the different $Q^2$ dependence of the involved distributions in the allowed range of $x$ values. As the transversity is unknown experimentally, this sensitivity to $Q^2$ is an important argument for future experiments. The present results suggest the possibility of measurable asymmetries at moderate values of $Q^2$ in the kinematic conditions of the proposed $\cal PAX$ experiment, thus obtaining direct access to the quark transversity distribution.

[Note added in the proof: During the revision process a phenomenological analysis of available data appeared~\cite{ABUKMPT07} and the transversity distributions for up and down quarks were shown to have opposite sign and a smaller size than their positivity bounds in agreement with the results of the present model.]

\medskip

\section*{Acknowledgements}

This research is part of the EU Integrated Infrastructure Initiative Hadronphysics Project under contract number RII3-CT-2004-506078. 




\clearpage



\begin{thebibliography}{99}

\bibitem{uno}
B.~Lampe and E.~Reya, Phys. Rep. 332, 1 (2000); \\
R.L.~Jaffe, hep-ph/9710465; \\
M.~Dittmar {\it et al.\/}, hep-ph/0511119.


\bibitem{Barone}
V.~Barone, A.~Drago, and P.G.~Ratcliffe, Phys. Rep. 359, 1 (2002); \\
V.~Barone and P.G.~Ratcliffe, {\it Transverse Spin Physics\/}, World Scientific, Singapore, 2003.

\bibitem{Ralston79}
J.P.~Ralston and D.~Soper, Nucl. Phys. B 152, 109 (1979).

\bibitem{Cortes92}
J.L.~Cortes, B.~Pire, and J.P.~Ralston, Z. Phys. C 55, 409 (1992).

\bibitem{Ji92}
Xiangdong Ji, Phys. Lett. B 284, 137 (1992).

\bibitem{JaffeJi}
R.L.~Jaffe and Xiangdong Ji, Phys. Rev. Lett. 67, 552 (1991); Nucl. Phys. B 375, 527 (1992).

\bibitem{Bunce00}
G.~Bunce, N.~Saito, J.~Soffer, and W.~Vogelsang, Ann. Rev. Nucl. Part. Sci. 50, 525 (2000).

\bibitem{PAX}
P.~Lenisa, F.~Rathmann {\it et al.\/} ($\cal PAX$ Collaboration), {\it Technical proposal for Antiproton--Proton Scattering Experiments with Polarization\/}, hep-ex/ 0505054.

\bibitem{EGS04}
A.V.~Efremov, K.~Goeke, and P.~Schweitzer, Eur. Phys. J. C 35, 207 (2004).

\bibitem{ABDN04}
M.~Anselmino, V.~Barone, A.~Drago, and N.N.~Nikolaev, Phys. Lett. B 594, 97 (2004).

\bibitem{BCCGR06}
V.~Barone, A.~Cafarella, C.~Corian\`o, M.~Guzzi, and P.G.~Ratcliffe, Phys. Lett. B 639, 483 (2006).

\bibitem{diehlbrodsky}
M.~Diehl, Th.~Feldmann, R.~Jakob, and P.~Kroll, Nucl. Phys. B 596, 33 (2001); \\
S.J.~Brodsky, M.~Diehl, and D.S.~Hwang, Nucl. Phys. B 596, 99 (2001).

\bibitem{BPT03}
 S. Boffi, B. Pasquini, and M. Traini, Nucl. Phys. B 649, 243 (2003).

\bibitem{BPT04}
S. Boffi, B. Pasquini, and M. Traini, Nucl. Phys. B 680, 147 (2004).

\bibitem{PPB05}
B.~Pasquini, M.~Pincetti, and S.~Boffi, Phys. Rev. D 72, 094029  (2005).

\bibitem{PB06}
B.~Pasquini and S.~Boffi, Phys. Rev. D 73, 094001 (2006).

\bibitem{BL}
S.J. Brodsky and G.P. Lepage, in: A.H. Mueller (ed.), {\it Perturbative Quantum
Chromodynamics}, World Scientific, Singapore, 1989, p.93; \\
S.J. Brodsky, H.-Ch. Pauli, and S.S. Pinsky, Phys. Rep. 301, 299 (1998).

\bibitem{MaSS}
I.~Schmidt and J.~Soffer, Phys. Lett. B 407, 331 (1997); \\
Bo-Qiang Ma, I.~Schmidt, and J.~Soffer, Phys. Lett. B 441, 461 (1998).

\bibitem{soffer}
J. Soffer, Phys. Rev. Lett. 74, 1292 (1995).

\bibitem{Schlumpf94a}
F.~Schlumpf, J. Phys. Nucl. Part. Phys. G 20, 237 (1994).

\bibitem{Cardarelli}
F.~Cardarelli and S.~Simula, Phys. Rev. C 62, 065201 (2000).

\bibitem{JaffeRoss}
R.L.~Jaffe and G.G.~Ross, Phys. Lett. B 93, 313 (1980).

\bibitem{DGLAP}
Yu.L.~Dokshitzer, JETP (Sov. Phys.) 46, 641 (1977) [Zh. Exsp. Teor. Fiz. 73, 1216 (1977)]; \\
V.N.~Gribov and L.N.~Lipatov, Sov. J. Nucl. Phys. 15, 438, 675 (1972) [Yad. Fiz. 15, 781, 1218 (1972)]; \\
L.N.~Lipatov, Sov. J. Nucl. Phys. 20, 94 (1975) [Yad. Fiz. 20, 181 (1974)]; \\
G.~Altarelli and G.~Parisi, Nucl. Phys. B 126, 298 (1977).

\bibitem{Baldracchini81}
F.~Baldracchini, N.S.~Craigie, V.~Roberto, and M.~Socolovsky, Fortschr. Phys. 29, 505 (1981).

\bibitem{Artru90}
X.~Artru and M.~Mekhfi, Z. Phys. C 45, 669 (1990).

\bibitem{Vogelsang93}
W.~Vogelsang and A.~Weber, Phys. Rev. D 48, 2073 (1993).

\bibitem{CKM94}
A.P.~Contogouris, B.~Kamal, and J.~Merebashvili, Phys. Lett. B 337, 169 (1994).

\bibitem{Kamal96}
B.~Kamal, Phys. Rev. D 53, 1142 (1996).

\bibitem{Hayashigaki97}
A.~Hayashigaki, Y.~Kanazawa, and Y.~Koike, Phys. Rev. D 56, 7350 (1997).

\bibitem{Kumano97}
S.~Kumano and M.~Miyama, Phys. Rev. D 56, R2504 (1997).

\bibitem{Vogelsang98}
W.~Vogelsang, Phys. Rev. D 57, 1886 (1998).

\bibitem{Barone97}
V.~Barone, Phys. Lett. B 409, 499 (1997).

\bibitem{Barone97b}
V.~Barone, T.~Calarco, and A.~Drago, Phys. Lett. B 390, 287 (1997).

\bibitem{HKM96b}
M.~Hirai, S.~Kumano, and M.~Miyama, Comp. Phys. Comm. 111, 150  (1996).

\bibitem{CC04}
A.~Cafarella and C.~Corian\`o, Comp. Phys. Comm. 160, 213  (2004).

\bibitem{Miyama96}
M.~ Miyama and S.~Kumano, Comp. Phys. Comm. 94, 185 (1996).

\bibitem{HKM96a}
M.~Hirai, S.~Kumano, and M.~ Miyama, Comp. Phys. Comm. 108, 35  (1996).

\bibitem{alpha}
T.~Weigl and W.~Melnitchouk, Nucl. Phys. B 465, 267 (1996); \\
M.~Traini, V.~Vento, A.~Mair, and A.~Zambarda, Nucl. Phys. A 614, 472 (1997); \\
B.~Pasquini, M.~Traini, and S.~Boffi, Phys. Rev. D 71, 034022 (2005).

\bibitem{GRV}
M.~Gl\"uck, E.~Reya, and A.~Vogt, Eur. Phys. J. C 5, 468 (1998).

\bibitem{GRSV}
M.~Gl\"uck, E.~Reya, M.~Stratmann, and W.~Vogelsang, Phys. Rev. D 63, 094005 (2001); \\
J.~Bl\"umlein and H.~B\"ottcher, Nucl. Phys. B 636, 225 (2002); \\
Y.~Goto {\it et al.\/}, Phys. Rev. D 62, 034017 (2000); Int. J. Mod. Phys. A 18, 1203 (2003).

\bibitem{MSSV98}
O.~Martin, A.~Sch\"afer, M.~Stratmann, and W.~Vogelsang, Phys. Rev. D 57, 3084 (1998).

\bibitem{MSSV99}
O.~Martin, A.~Sch\"afer, M.~Stratmann, and W.~Vogelsang, Phys. Rev. D 60, 117502 (1999).

\bibitem{BouS94}
C.~Bourrely and J.~Soffer, Nucl. Phys. B 423, 329 (1994).

\bibitem{BS95}
C.~Bourrely and J.~Soffer, Nucl. Phys. B 445, 341 (1995).

\bibitem{Barone97a}
V.~Barone, T.~Calarco, and A.~Drago, Phys. Rev. D 56, 527 (1997).

\bibitem{BR05}
A.~Bianconi and M.~Radici, Phys. Rev. D 72, 074013 (2005).

\bibitem{Shimizu}
H.~Shimizu, G.~Sterman, W.~Vogelsang, and H.~Yokoya, Phys. Rev. D 71, 114007 (2005).
 
\bibitem{ABUKMPT07}
M.~Anselmino, M.~Boglione, U.~D'Alesio, A.~Kotzinian, F.~Murgia, A.~Prokudin, and C.~Turk, Phys. Rev. D 75, 054032 (2007).


\end{thebibliography}
\end{document}